\documentclass[twocolumn,showpacs,preprintnumbers,amsmath,amssymb]{revtex4}

\usepackage{graphicx}% Include figure files
\usepackage{dcolumn}% Align table columns on decimal point
\usepackage{bm}% bold math

\begin{document}

\title{Adiabatic and isothermal sound waves: the case of supercritical nitrogen}

\author{F. Bencivenga$^{1}$, A. Cunsolo$^{2}$, M. Krisch$^{1}$, G. Monaco$^{1}$,
G. Ruocco$^{3}$ and F. Sette$^{1}$\\
$^{1}$ European Synchrotron Radiation Facility. B.P. 220 F-38043
Grenoble, Cedex France.\\
$^{2}$ CRS SOFT-INFM-CNR - Operative Group in Grenoble c/o ILL,
B.P. 220 F-38043 Grenoble, Cedex France. \\
$^{3}$ Dipartimento di Fisica and CRS SOFT-INFM-CNR,
Universit\'{a} di Roma
"La Sapienza", Roma, Italy.\\
 }

\date{\today}

\begin{abstract}

The acoustic sound dispersion of nitrogen in its liquid and
supercritical phases has been studied by Inelastic X-Ray
Scattering. Approaching supercritical conditions, the gradual
disappearance of the positive sound dispersion, characteristic of
the low temperature liquid, is observed. In the supercritical
state, evidence for a crossover between adiabatic and isothermal
sound propagation regimes is inferred by an analysis of the
dynamic structure factor based on generalized hydrodynamics.

\end{abstract}

\maketitle

The development of Inelastic X-ray Scattering ($IXS $)
\cite{IXS,Sette,FraGiaLong,Burk,tullrev} has opened up new
opportunities in the study of the high-frequency dynamics of
liquids at nanometer distances and picosecond timescales. Various
$IXS$ investigations unveiled that, in the proper thermodynamical
conditions, $structural$ relaxation processes occur in the THz
frequency range. These processes manifest themselves by an upward
bending of the sound dispersion relation, which deviate from the
low-frequency linear behavior dictated by the adiabatic sound
velocity. Accordingly, the sound velocity increases from the
adiabatic value, $c_{s}$, to the "infinite" frequency value,
$c_{\infty}$
\cite{Sette,FraGiaLong,Burk,tullrev,neon,watPRE,lastPRE,PRLMK,ammonia,balu1,balu2,Amon,tull1,tull2,tull3,lowtwat,simuwat}.
If $\Omega _{L}(Q)$ is the frequency of the probed acoustic
excitation with wave vector $Q$, and $\tau_\alpha$ is the
characteristic relaxation time of the structural relaxation
process, the system exhibits a liquid-like ($viscous$) behavior if
the condition $\Omega _{L} (Q)\tau_\alpha \ll 1$ is fulfilled: in
this limit the dispersion relation is $\Omega _{L}(Q)=c_{s}Q$. In
the opposite limit, the $elastic$ one, where $\Omega _{L}(Q)
\tau_\alpha \gg 1$, structural rearrangements are too slow to
efficiently dissipate the energy of the acoustic waves: in this
limit $\Omega _{L}(Q)=c_{\infty} Q$. A schematic picture of this
$visco-elastic$ variation of the sound velocity form
$c_{s}\rightarrow c_{\infty}$ is shown in Fig. 1a (full thick
line), where the cross-over condition $\Omega
_{L}(Q)=1/\tau_\alpha$ is indicated by the vertical arrow, and the
adiabatic and infinite linear dispersion relations are indicated
by the the full and dashed lines respectively. Here the dash-dot
line represents the value of $1/\tau_\alpha$, which is here
assumed to be $Q$-independent.

\begin{figure}[htbp]
\begin{center}
\includegraphics[width=7cm]{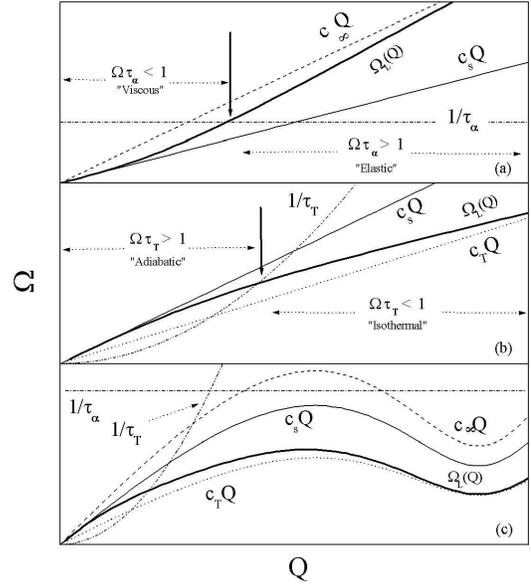}
\end{center}
\caption{\emph{Schematic representation of the various sound
regimes highlighting the two cross-over conditions. Upper panel:
viscoelastic transition taking place at $\Omega_{L}=1/\tau$;
middle panel: isothermal transition taking place at
$\Omega_{L}=D_{T}Q^{2}$; bottom panel: realistic non linear
dispersions under favorable experimental conditions for
$\Omega_{L} \ll 1/\tau$ and $\Omega_{L}\approx D_{T}Q^{2}$.
Vertical arrows indicate the crossover regions as discussed in the
text. The isothermal (dotted line), adiabatic (solid line),
infinite frequency (dashed line) and apparent (thick solid line)
sound dispersions are reported together with the relaxation
frequency $1/\tau$ (dashed-dotted line) and thermal frequency
$D_{T}Q^{2}$ (dashed-double dotted line).}}
\end{figure}

The structural relaxation process, in principle, is not the only
relaxation process that could affect the high frequency dynamics
of a liquid. On general grounds, also other physical processes, as
for example thermal diffusion, can modify the behavior of sound
waves. In this case, given a thermal diffusion coefficient $D_{T}$
and a thermal relaxation time $\tau_T= 1/ D_{T}Q^{2}$, a
modification of the sound velocity takes place around a cross-over
frequency given by $\Omega_L(Q) \approx D_{T}Q^{2} $. In this
case, for $D_{T}Q^{2} \ll \Omega_L(Q)$ the thermal diffusion is
much slower than the period of acoustic waves, which therefore
propagate adiabatically, i.e. without any thermal exchange with
the local environment. In the opposite limit, $D_{T}Q^{2}\gg
\Omega_L(Q) $, the thermalization of the acoustic wave is
instantaneous, and the effective propagation mechanism becomes
isothermal with sound velocity  $c_{T}=\gamma ^{-1/2}c_{s}$
($\gamma $ is the specific heat ratio). One therefore expects to
observe a transition between the adiabatic and the isothermal
regimes when $\Omega_{L}(Q)= D_{T}Q^{2}$. This case is
schematically presented in Figure 1b, where linear lines represent
the adiabatic (full line) and isothermal (dotted line) sound
dispersion, and the dash-dot-dot line indicates the $Q$ dependence
of $1/\tau_T(Q)$.

Contrary to the visco-elastic transition, which has been
experimentally documented in many liquids
\cite{FraGiaLong,tullrev,neon,watPRE,lastPRE,PRLMK,ammonia,balu1,balu2,Amon,tull1,tull2,tull3,lowtwat,simuwat},
no firm evidence of the adiabatic to isothermal transition is
available. This is mainly due to the following reasons: i) The
amplitude of the dispersive effect between $c_s$ and $c_T$ is
proportional to $\gamma-1$ which, in most liquids, is close to
zero. ii) The effect is often completely masked by the competing
visco-elastic transition. iii) The crossover occurs typically at
large $Q$, often above the position of the first sharp diffraction
peak (FSDP) of the static structure factor, $S(Q)$. Here, however,
the finite-$Q$ generalization of the isothermal sound speed
$c_T(Q)$ - which is proportional to $ 1/S(Q)^{1/2}$ -
\cite{textbook}, becomes too small to be reliably extracted from
the measured $S(Q,\omega)$ spectra.

These obstacles to the observation of the adiabatic to isothermal
sound propagation transition can be reduced close to supercritical
conditions for the following reasons: i) the competing positive
dispersion induced by the viscoelastic transition is expected to
be much weaker. ii) The crossover condition for the viscoelastic
transition is expected to move at higher $Q$, i.e. higher
frequency, since the structural rearrangements occur on shorter
time scales. iii) Moreover, the intensity of the FSDP is much
weaker, and consequently $ c_{T}(Q) $ does not display any longer
a pronounced minimum in the vicinity of the FSDP. iv) Finally, the
sound speed decreases on approaching the critical point, and
therefore the transition will move towards smaller $Q$ values.
This scenario is schematically pictured in Figure 1c, where the
combined effects of the structural and thermal relaxation
processes are model within the framework of the molecular
hydrodynamics \cite{textbook}. The values of $\tau_\alpha$ and
$\tau_T(Q)$ are chosen such that the transition is actually
determined by the thermal relaxation process, as indicated by the
arrow.

This letter, guided by the above considerations, reports an IXS
study of liquid and supercritical nitrogen. Nitrogen has been
chosen because the two thermodynamic regimes (liquid and
supercritical) can be easily obtained within an experimentally
accessible pressure and temperature range, and because of its
favorable inelastic scattering cross section. Figure 2 shows the
portion of the explored thermodynamic plane. The inset reports the
expected Q-values, $Q^{*}(T)$, for which the crossover from the
adiabatic to the isothermal regime should occur:
$Q^{*}(T)=c_{s}/D_{T}$. The values of $c_{s}$ and $D_{T}$ were
obtained from the nitrogen Equation-of-State (EoS) \cite{NIST}.

\begin{figure}[htbp]
\begin{center}
\includegraphics[width=7cm]{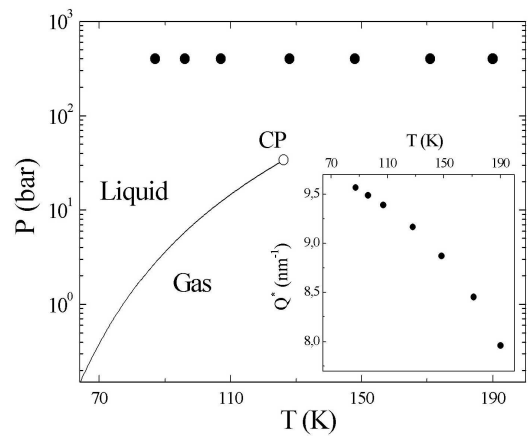}
\end{center}
\caption{\emph{Phase diagram of nitrogen. The investigated
thermodynamic points are reported as full circles. The solid line
represents the coexistence curve, ending at the critical point CP
($T_{c}$=126.2 K, $P_{c}=34$ bar,  $\rho_{c}=0.313$ g/cm3). In the
inset we report the expected Q-values for the isothermal
transition for the sound propagation, as estimated from the
nitrogen equation-of-state using $Q=c_{s}/D_{T}$.}}
\end{figure}

The experiment was performed on ID28 at the European Synchrotron
Radiation Facility (ESRF). The beamline was set up with an
instrumental resolution function of $1.5 meV$
full-width-half-maximum (FWHM). The sample was embedded in a large
volume, high pressure cell kept in thermal contact with the cold
finger of a cryostat. The optical windows were two 1 mm thick
diamonds disks, and the sample length amounted to 10 mm. The
pressure stability was better than 5 bar over the acquisition time
of a typical spectrum. IXS spectra were recorded following an
isobaric path at 400 bar from subcritical to supercritical
conditions (full dots in figure 2). Cell contribution as
determined by empty cell measurement yielded a negligible
contribution to the signal for all exploited scattering
geometries. Multiple scattering has been estimated to be
negligible. The recorded signal is therefore proportional to the
dynamic structure factor, $S(Q,\omega)$, convoluted with the
instrumental resolution function.

\begin{figure}[htbp]
\begin{center}
\includegraphics[width=7.5cm]{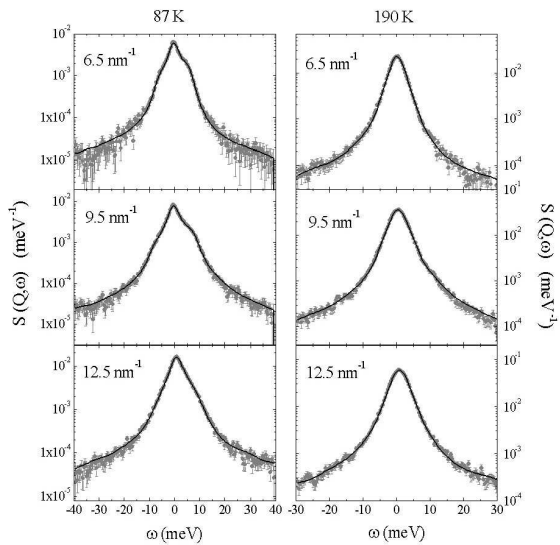}
\end{center}
\caption{\emph{Selection of IXS spectra of liquid nitrogen at
$P=400$ bar and two temperatures (87 K and 190 K) at momentum
transfers as indicated in the individual panels. The experimental
data (full circles) are reported together with their best fit
lineshapes (solid lines).}}
\end{figure}

The data analysis has been performed using a line-shape model
derived within the framework of the memory function formalism
\cite{textbook}. In this model the dynamic structure factor is
written as:

\begin{equation} \label{sqw}
S(Q,\omega )=\frac{\hbar \omega }{K_{B}T}\left[ n(\omega
)+1\right] Im[\omega ^{2}-q^{2}c_{T}^{2}-i\omega m_{Q}(\omega
)]^{-1}
\end{equation}

where $n(\omega)$ is the Bose factor and $m_{Q}(\omega )$ is the
Fourier transform of the time dependent memory function
\cite{textbook,watPRE,lastPRE}, which has been approximated by:

\begin{eqnarray}
\label{memory}
m_{Q}(t)=c_{T}^{2}(Q)[\gamma(Q)-1]Q^{2}e^{-D_{T}(Q)Q^{2}t}+ \\
+[c_{\infty}^{2}(Q)-c_{s}^{2}(Q)]Q^{2}e^{-t/\tau_\alpha(Q)}+\Delta^2(Q)
\delta (t)\nonumber
\end{eqnarray}

Following the prescriptions of molecular hydrodynamics, all the
thermodynamic quantities entering Eq.~\ref{memory} are assumed as
$Q$ dependent extensions of their respective macroscopic
counterparts.

The $\delta$ function in Eq.\ref{memory} accounts for the fast
dynamics which induces an instantaneous time decay of $m_{Q}(t)$
\cite{watPRE,lastPRE}. The second exponential term of
Eq.\ref{memory} accounts for the $visco-elastic$ transition that
disperses the sound velocity, with increasing $Q$, from its low to
its high frequency limiting value, $c_{s}(Q)$ and $c_{\infty}(Q)$
respectively. Finally, the first term in Eq.~\ref{memory} accounts
for the thermal diffusion process. In the low-$Q$ limit the
$thermal$ part of the memory function yields a central peak in the
spectral density, whose integrated intensity is proportional to
$\gamma(Q)-1$, and its half-width is given by $D_{T}(Q)Q^{2}$.
Consequently the proposed model predicts the "classical"
hydrodynamics description at low $Q$.

The experimental data are described by the convolution of the
model reported in Eq.s \ref{sqw} and \ref{memory} with the
experimentally determined instrument resolution function. In the
fitting procedure, the parameters $\tau_\alpha(Q)$, $c_{\infty
}(Q)$, $\Delta^2(Q) $ and an overall intensity factor $A$ are left
as free parameters. Owing to the lack of experimental or
computational results on the $Q$-dependence of $\gamma $ and
$D_{T}$, these variables have been fixed to their thermodynamic
($Q\rightarrow 0 $) values, as derived from the EoS of nitrogen
\cite{NIST}. The parameters $c_{T}(Q) $ and $c_{s}(Q) $ have been
fixed to the values calculated through \cite{textbook}:

\begin{equation} \label{sound}
c_{T}(Q)=\gamma
^{-1/2}c_{s}(Q)=[\frac{K_{B}T}{MS(Q)}]^{\frac{1}{2}}
\end{equation}

where $K_{B}$ and $ M $ are the Boltzmann constant and the
molecular mass of nitrogen, respectively. The static structure
factor S(Q) was determined experimentally by recording the
energy-integrated scattering, corrected for the known nitrogen
form factor, $f(Q)$ \cite{form}, as well as for all geometrical
artifacts. In order to put $S(Q)$ on an absolute scale, we have
normalized the measured intensity to fulfill the compressibility
constraint in the $Q=0$ limit. Finally, the apparent sound
velocity of acoustic excitations, $c_{L}(Q)$, has been determined
using the relation $c_{L}(Q) =\Omega _{L}(Q)/Q $, where
$\Omega_{L}(Q)$ is the frequency value of the maximum of the
longitudinal current spectra, $\omega ^{2}S(Q,\omega)/Q^{2}$.

Selected IXS spectra and their best fits are shown in Fig.3. Each
spectrum typically covers an energy transfer range of $\pm 40$ meV
and was collected in the 2 to 14 $nm^{-1}$ $Q$ range. The
logarithmic plot emphasizes the overall good agreement between
experimental and model line shapes, even in the tails of the
spectra.

The comparisons between the apparent dispersion relation,
$\Omega_{L}(Q)$ $vs.$ $Q$ (open circles), and both isothermal
(dotted line) and adiabatic (full line) ones, are shown in Fig. 4
at the four different temperatures 87, 128, 171, and 191 K. In the
same figure we also report the infinite frequency dispersion
relation (full dots) and the inverse of the structural relaxation
time $1/\tau_\alpha(Q)$ (dashed-dot line) as derived from the best
fit result. The inverse of the thermal relaxation time,
$D_{T}Q^{2}$, is also reported as derived from the nitrogen EoS
(dash-dot-dot line). The results of Fig.4a, corresponding to the
lowest investigated temperature (87 K), show that the apparent
dispersion is systematically higher than the adiabatic one.
Although the $c_{\infty}$-limit is not completely reached, in this
thermodynamic point the high frequency dynamics of the liquid is
mostly influenced by the structural relaxation process. This
influence is considerably weaker at $T=128$ K (Panel b) due to the
higher value of $1/\tau_\alpha(Q)$ which prevents even-more to
reach the viscoelastic crossover conditions. On further increase
of the temperature (Panel c and d), $1/\tau_\alpha(Q)$ moves
definitely out of the probed frequency range and any hint of
positive dispersion on the sound velocity disappears. In fact,
with increasing temperature the apparent dispersion relation
initially approaches the adiabatic sound dispersion, but at the
highest temperature is observed to go below the adiabatic
dispersion and to approach the isothermal one: this is the onset
of the expected transition between adiabatic and isothermal
regimes. This effect is most clearly seen for the three highest
$Q$ points between $9.5$ and $14$ $nm^{-1}$.

\begin{figure}[htbp]
\begin{center}
\includegraphics[width=7.5cm]{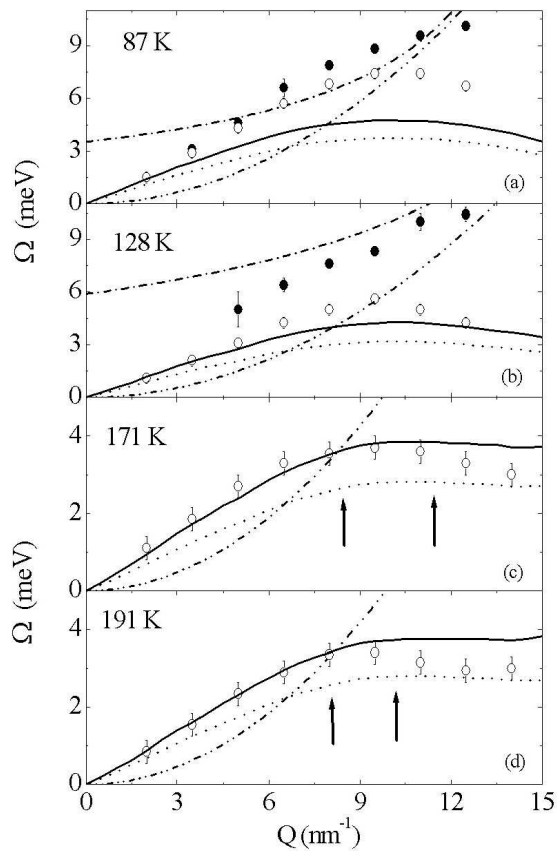}
\end{center}
\caption{\emph{Dispersion curves of the longitudinal acoustic mode
of nitrogen at 87, 128, 171 and 190 K. The full and dotted lines
represent the adiabatic and isothermal dispersions, derived from
$S(Q)$ measurements, as discussed in the text; the open circles
indicate the maxima of longitudinal current spectra. The
dash-dotted and dash-double dotted lines represent $1/\tau (Q)$
and $D_{T}Q^2$, respectively; the former has been evaluated
interpolating the results of the performed lineshape analysis,
which allowed also to extract the values of infinite-frequency
sound dispersion, reported as full circles.}}
\end{figure}

The arrows in Figure 4 (panels c and d) indicate the $Q$ points
where one would expect the adiabatic-isothermal transition
($\Omega_L(Q)=D_T Q^2$, left arrows) and where the transition
actually occurs (right arrows): a $Q$ value definitively higher
than the one estimated by the left arrow. This difference between
expected and observed transition $Q$ points is an indication of
the $Q$-dependence of the parameter $D_{T}$, which we neglected so
far. This behavior is in agreement with the expectations suggested
by numerical simulations of various molecular liquids
\cite{DTwat,Li4Pb} pointing to a monotonic decrease of $D_{T}(Q)$
with increasing $Q$. As a consequence, the position of the
crossover is shifted towards higher momentum transfers. Even
though the transition to the isothermal regime is not fully
completed, its occurrence is inferred from this systematic trend.
Furthermore, it is worthwhile recalling that the present results
have been obtained without imposing any constraint on the line
shape fitting parameters. Moreover, the values of the isothermal
sound dispersions have been determined independently through the
measured $S(Q)$.

In conclusion our observations can be summarized as follows: i)
the apparent sound branch of nitrogen show clear signatures of a
positive sound dispersion in the liquid phase, thus witnessing the
presence of a structural relaxation process. ii) The positive
sound dispersion gradually disappears while reaching supercritical
conditions, most likely due to the shift of $1/\tau_\alpha(Q)$
above the $\Omega_{L}(Q)$ values. iii) At the highest investigated
temperatures, the transition from the adiabatic to the isothermal
sound propagation is observed.

These findings indicate that, in the liquid phase, the high
frequency dynamic of a simple, non-associated fluid, such as
nitrogen, is mainly ruled by the structural relaxation, as
manifested by the positive sound dispersion. In contrast, in the
supercritical phase, the role of the structural relaxation becomes
negligible and the most evident dispersive effect is the bending
down of the apparent dispersion, from its adiabatic value to the
isothermal one. In this regime, the thermal relaxation process
rules the high frequency dynamics.

\section{Acknowledgments}

We are grateful to Dr. T. Scopigno for critical reading of the
manuscript. We are also grateful to D. Gambetti and R. Verbeni for
help in the preparation of the experiment.

\end{document}